\documentstyle[sprocl]{article}

\def\discrete{Z_2}
\def\oneht{\textstyle{1\over 2} }
\def\sss{\scriptscriptstyle}
\def\ql{{ Q_{\sss L} }}
\def\qr{{ Q_{\sss R} }}
\def\pl{{ {P}_{\sss L} }}
\def\pr{{ {P}_{\sss R} }}
\def\parlam{^{\scriptstyle\lambda}}

\def\be{\begin{equation}}
\def\ee{\end{equation}}
\def\bea{\begin{eqnarray}}
\def\eea{\end{eqnarray}}
\begin{document}

%{\small \noindent{U.Md. PP\# 99-}\hfill{hep-ph/9810298}}
%\vspace{0.2in}

{\small \noindent{}\hfill{hep-ph/9810298}}
\vspace{0.2in}

\title{THE ADVANTAGES OF BEING VECTORLIKE}

\author{SILAS R.~BEANE}

\address{Department of Physics, University of Maryland \\
College Park, MD 20742-4111, USA \\E-mail: sbeane@physics.umd.edu} 

\maketitle\abstracts{We construct an asymptotically free gauge theory
with an $SU(2)\times SU(2)\times U(1)$ vectorlike flavor symmetry. We
show that this gauge theory has properties which provide insight into
aspects of low-energy QCD.}

\section{Introduction}

Consider an asymptotically free gauge theory with global symmetry
group $G$. At short distances this gauge theory allows a perturbative
definition. One fundamental question at long distances relates to how
the symmetry group $G$ is realized in the vacuum.  Virtually no
progress has been made in answering this question from first
principles in QCD, the gauge theory responsible for all hadronic and
nuclear phenomena. One might hope that ultimately lattice methods will
provide insight. Unfortunately, there is at present no lattice
formulation of gauge theory which preserves $G$ if $G$ is chiral, and
therefore at present the lattice cannot reveal how QCD flavor
symmetries are realized in the vacuum.

For purposes of this paper, QCD has two massless flavors and global
symmetry group $G=SU(2)\times SU(2)\times U(1)$.  This global symmetry
group can be decomposed into $H=SU(2)_V \times U(1)$ which corresponds
to isospin and baryon number and is {\it vectorlike} since it does not
imply massless quarks, and $SU(2)_A$ which is {\it chiral} and
therefore exact only in the limit of vanishing quark masses. No-go
theorems formulated for QCD-like gauge theories imply that vectorlike
symmetries are not spontaneously broken~\cite{vafa}. Therefore, on
theoretical grounds we expect that in the low-energy theory $H$ will
be unbroken, and $SU(2)_A$ can be unbroken, completely broken or
partially broken by the vacuum. Observation reveals that $SU(2)_A$ is
broken completely in the vacuum.

In this paper we will motivate study of other gauge theories with QCD
flavor symmetries by considering general properties of the nonlinear
realization of $G$. We will show that the manifold possible $G$
transformation properties of the hadronic mass-squared matrix suggest
the existence of other gauge theories with precisely the same symmetry
breaking pattern as QCD.  We will construct an asymptotically free
gauge theory with a $G=SU(2)\times SU(2)\times U(1)$ flavor
symmetry. This gauge theory is different from QCD both in its matter
content and in that the entire global symmetry group is
vectorlike. Therefore the masses of the quarks in this theory are
unconstrained by symmetries. We will show that this gauge theory has
interesting properties which provide insight into several mysterious
aspects of low-energy QCD.

\section{The Two Faces of the Nonlinear Realization}
\subsection{Chiral Perturbation Theory}\label{subsec:cpt}

Consider an underlying theory with flavor symmetry group
$G=SU(2)\times SU(2)\times U(1)$ which is spontaneously broken to
$H=SU(2)_V \times U(1)$. The assumption that $SU(2)_A$ is broken by
the vacuum leads to a wealth of information about the low-energy
theory, which is true for all underlying theories with the same
pattern of symmetry breaking~\cite{ccwz}.  There are $3$ Goldstone
bosons, identified with pions.  Using the theory of nonlinear
realizations it is straightforward to construct the most general
low-energy lagrangian of pions living on the coset space
$G/H=SU(2)\times SU(2)/SU(2)$. The basic physics underlying this
construction is simple. Since it costs no energy to move from one
point on the vacuum manifold to another, pions couple to themselves
and other hadrons only through derivative interactions. Hence the
assumed pattern of symmetry breaking implies that scattering
amplitudes involving pions are expansions in powers of pion energy.
The parameters that appear in this expansion are unconstrained by the
pattern of symmetry breaking, but can be fit to one experiment in
order to predict another.  This method is known as chiral perturbation
theory.

\subsection{Asymptotic Matching}\label{subsec:am}

Chiral perturbation theory is not the only information encoded in the
nonlinear realization.  The scattering amplitudes involving pions are
polynomials in energy.  Therefore, in order to get reasonable
asymptotic behavior, the axial couplings of the hadrons must be
related in special ways. This leads to an apparent paradox. Presumably
couplings can be related only by symmetry. However, $G$ is the only
symmetry in the problem and $G$ is spontaneously broken!  What then
relates the axial couplings?

Long ago Weinberg~\cite{alg}, and independently Wess and
Zumino~\cite{zumi}, answered this question by considering pion-hadron
scattering in helicity conserving (collinear) Lorentz frames. The
fundamental observation is that the nonlinear realization also
constrains the form of scattering amplitudes involving any number of
pions and hadrons expanded in {\it inverse} powers of energy. Consider
the elastic scattering process $\pi\alpha\rightarrow\pi\beta$ where
$\alpha$ and $\beta$ are any single-hadron states. It is possible to
extract the coefficient of the scattering amplitude that scales as the
zeroth power of energy. The crossing-odd and -even amplitudes are,
respectively,

\begin{eqnarray}
{{\cal M}^{(-)\;\lambda}_{\beta b,\alpha a}}(\omega )&\propto&
\lbrace i\epsilon _{abc}T_{c}- 
\lbrack {X_{a}\parlam},\, {X_{b}\parlam}\rbrack
\rbrace _{\beta\alpha}{\,{\omega^{\sss 0}}}  + O(1/{\omega} )+O(\omega ) 
\label{eq:matchcondsa} \\
{{\cal M}^{(+)\;\lambda}_{\beta b,\alpha a}}(\omega )&\propto&
\lbrack {X_{b}\parlam},\,\lbrack { M^{2}},\,{X_{a}\parlam}
\rbrack\rbrack  _{\beta\alpha}{\,{\omega^{\sss 0}}}+ O(1/{\omega} )+O(\omega )
\label{eq:matchcondsb}
\end{eqnarray}
where ${T_a}$ is the isospin matrix, and ${X_a\parlam}$ is an axial vector
coupling matrix, related to the matrix element of the process ${\alpha
(p,\lambda )}\rightarrow{\beta (p',\lambda')}+{\pi}{(q,a)}$ in any
frame in which the momenta are {\it collinear}, and $\lambda$ is
helicity ---which is conserved in the collinear frame.  The
coefficients of positive powers of energy always contain counterterms
which can be assigned any value and so are uninteresting. It is the
coefficients of the zeroth power of $\omega$ which are interesting,
since they involve only the axial vector coupling matrix and the
hadronic mass-squared matrix in algebraic form.

The values taken by these coefficients determine how the low-energy
theory matches to the underlying theory. Since ${SU(2)}\times{SU(2)}$
is a symmetry of the underlying theory we have the matching condition
${\cal M}^{{(-)}}=0$. Together with Eq.~(\ref{eq:matchcondsa}), the
defining relations, $\lbrack
T_{a},\,T_{b}\rbrack=i\epsilon_{abc}T_{c}$ and $\lbrack
T_{a},\,{{X_{b}\parlam}}\rbrack= 
i\epsilon _{abc}{{X_{c}\parlam}}$ close the chiral
algebra and so {\it for each helicity}, $\lambda$, hadrons fall into
representations of ${SU(2)}\times{SU(2)}$, {\it in spite of the fact
that the group is spontaneously broken}. This statement of chiral
symmetry is simply a generalization of the Adler-Weisberger sum
rule. The matching condition ${\cal M}^{{(-)}}=0$ is equivalent to the
statement that the crossing-odd amplitude satisfies an unsubtracted
dispersion relation~\cite{alg}.

The matching condition for the crossing-even amplitude is especially
interesting since it determines the asymptotic behavior of the total
cross-section.  It is clear from Eq.~(\ref{eq:matchcondsb}) that the
matching condition is determined by the transformation property with
respect to $SU(2)\times SU(2)$ of the hadronic mass-squared
matrix~\cite{alg}.  In general one can write

\begin{equation}
{{\hat M}^{2}}={\sum_{\cal R}} {{\hat M}^{2}_{\cal R}}
\label{eq:massreps}
\end{equation}
where ${\cal R}$ is a representation of $SU(2)\times SU(2)$. For
instance, if ${\cal M}^{{(+)}}=0$ then ${{\hat M}^{2}}$ transforms as
the singlet representation, ${\cal R}=(1,1)$.  As we will show, the
correct symmetry property of the mass-squared matrix is difficult to
infer from the QCD lagrangian because symmetry breaking is inherently
nonperturbative.

\section{Asymptotic Matching in QCD}

Can we learn anything about the transformation properties of ${{\hat
M}^{2}}$ directly from the QCD lagrangian?  In two-flavor massless QCD
the degrees of freedom relevant to $G$ are the Weyl fermions $Q_{\sss
L}$ and $Q_{\sss R}$ which transform as $(2,1)$ and $(1,2)$ with
respect to $SU(2)\times SU(2)$.  Consider a generic mass term $M{\bar
Q}Q$ which might represent a current quark mass or a constituent mass
induced by a $<{{\bar Q}Q}>$ condensate (equivalently we could
consider a sigma model).  We assign $M$ spurion transformation
property $(2,2)$ with respect to $SU(2)\times SU(2)$. In order to make
contact with the results of the previous section formulated in
helicity conserving Lorentz frames we can express this operator in
light-front coordinates which gives ${M^2}{Q_{\sss +}^\dagger}{Q_{\sss
+}}$~\cite{beane}. Here ${M^2}$ transforms as $(1,1)$.  This then
implies that ${{\hat M}^{2}}$ trasforms like a singlet giving the
matching condition ${\cal M}^{{(+)}}=0$, and total hadronic cross
sections should fall off rapidly asymptotically. This is of course in
violent disagreement with experiment. However, there is no
contradiction. The QCD lagrangian does not reveal the correct
transformation properties of the mass-squared matrix because anything
we infer from the QCD lagrangian and its degrees of freedom is tied to
perturbation theory, and the correct transformation properties are
nonperturbative. Presumably if there was a means of defining QCD at
long distance that respected $G$, the correct matching condition would
be manifest.

Weinberg~\cite{mended} made the ansatz ${{\hat M}^{2}}={{\hat
M}^{2}_{\sss (1,1)}}+{{\hat M}^{2}_{\sss (2,2)}}$ (${\cal
R}=(1,1)\oplus(2,2)$) which is equivalent to a Regge inspired
superconvergent sum rule and consistent with the observed
(approximately) constant behavior of total cross-sections; that is,
${{\cal M}^{(+)}_{ab}}\propto{\delta_{ab}}$.  He further showed that
the assumption $[{{\hat M}^{2}_{\sss (1,1)}},{{\hat M}^{2}_{\sss
(2,2)}}]=0$ fixes the reducible representations filled out by mesons
in the low-energy theory. This matching condition, ${{\cal
M}^{(+)}_{\alpha\beta}}\propto{\delta_{\alpha\beta}}$, is equivalent
to another Regge inspired superconvergent sum rule.  These assumptions
lead to remarkably accurate predictions in the low-energy theory, and
yet are mysterious from the QCD point of
view~\cite{alg}$^{\!,\,}$\,\cite{beane}$^{\!,\,}$\,\cite{mended}.

\section{Gauge Theory with Vectorlike $G$}
\subsection{Basic Formalism}\label{subsec:form}

Given the difficulty in understanding the transformation properties of
the hadronic mass-squared matrix from the QCD lagrangian, it is
natural to ask whether it is possible to construct other underlying
gauge theories with the same flavor symmetries as QCD but with a
mass-squared matrix whose transformation properties are nontrivial
and manifest. Surprisingly, the answer is yes.

In order to construct an asymptotically free gauge theory with {\it
vectorlike} $G$ flavor symmetry and all matter transforming
nontrivially with respect to the gauge group we require at least four
flavors of quarks, isodoublets $Q$ and $P$, transforming in the
fundamental representation of the gauge group, and one $G$
four-vector, Lorentz scalar, $\cal M$, transforming in the adjoint
representation of the gauge group.  The quantum number assignments are
as in Table~\ref{tab:exp}.

\begin{table}[t]
\caption{The $SU(2)\times SU(2)\times U(1)$ (global) and $SU(N)$ (gauge)
charge assignments.\label{tab:exp}}
\vspace{0.2cm}
\begin{center}
%\footnotesize
\begin{tabular}{|c|c|c|c|l|}
\hline
{} &{$SU(N)$} &{$SU(2)_L$} &{$SU(2)_R$} &{$U(1)_B$}  \\
\hline
{$\ql$, $\pr$} &{$\bf N$} &{$\bf{2}$} &{$\bf{1}$} 
&{${\textstyle {1/N}}$}  \\
\hline
{$\qr$, $\pl$} &{$\bf N$} &{$\bf{1}$} &{$\bf{2}$} 
&{${\textstyle {1/N}}$}  \\
\hline
{${\cal M}$} &{$\bf{{N^2} -1}$}  &{$\bf{2}$} 
&{$\bf{2}$} &{$0$}  \\
\hline
\end{tabular}
\end{center}
\end{table}

We have left the number of colors, $N$, of the gauge theory arbitrary.
The most general renormalizable lagrangian invariant with respect to
the flavor symmetries and $P$, $C$ and $T$ is:

\begin{eqnarray}
{\cal L}&=&\,{\bar \psi}i{\not \!\!{D}} \psi 
-{M_0}{\bar \psi}{\sigma _{\sss 1}}\psi 
+{\kappa_{\sss 1}}{\bar \psi}(A +i B{\gamma_5}{\sigma _{\sss 3}})\psi 
+{\kappa_{\sss 2}}{\bar \psi}({\sigma _{\sss 3}}A +i B{\gamma_5})\psi 
\nonumber \\
&&\, +{\oneht}\,{\it tr}(D_\mu {\cal M} D^\mu {{\cal M}^\dagger})
-{\textstyle {1\over 2}}\,{\it tr}(F_{\mu\nu}F^{\mu\nu})
-V({\cal M}{{\cal M}^\dagger})
\label{eq:gaugelag}
\end{eqnarray}
where $\psi=({Q}\;{P})^{\sss T}$ and ${\cal M}\equiv A+iB$. The trace
is over gauge and flavor quantum numbers and the $\sigma_{\sss i}$ are
Pauli matrices acting in the $\psi$ space.  This lagrangian admits a
$G=SU(2)\times SU(2)\times U(1)$ flavor symmetry and defines the gauge
theory for weak couplings. We have chosen $Q$ and $P$ to be of like
Parity.  It is easy to check that the gauge coupling is asymptotically
free for all $N$. The low-energy theory is therefore expected to be in
the confined phase.  The gauge theory is by construction anomaly free.
Note that if ${\kappa_{\sss 1}}={\kappa_{\sss 2}}=0$ the free fermion
theory is $U(4)$ invariant.  Therefore, the Yukawa operators break
$U(4)$ explicitly to $SU(2)\times SU(2)\times U(1)$.  There are
regions of parameter space where there is enhanced symmetry. When
${\kappa_{\sss 2}}=0$ there is a $\discrete$ symmetry that cannot be
constructed from any $G$ subgroup: $\psi\rightarrow {\sigma _{\sss
1}}\psi$, $B\rightarrow -B$.

\subsection{Vectorlike Asymptotic Matching}\label{subsec:am2}

Assume that $G=SU(2)\times SU(2)\times U(1)$ is broken spontaneously
to $H=SU(2)_V\times U(1)$ by the condensates $<{{\bar\psi}\psi}>$ and
$<{{\bar\psi}{\sigma _{\sss 3}}\psi}>$, where the latter condensate is
$\discrete$ violating. This is consistent with the observed pattern of
symmetry breaking in QCD. Note that due to the presence of the
fundamental scalars the Vafa-Witten theorem does not constrain
symmetry breaking in this gauge theory~\cite{vafa}.

Since the pattern of symmetry breaking is, by assumption, the same as
QCD, the low-energy theory is of the same form, as given by the
nonlinear realization. Of course in chiral perturbation theory the
values of the low-energy constants for the two underlying theories can
be different. What about asymptotic matching?  Here we can do better
than QCD. Of course we continue to have the matching condition ${\cal
M}^{{(-)}}=0$. But here we can also determine the crossing-even
matching condition.  Allowing constituent quark mass terms induced by
the condensates gives ${\hat M}={{\hat M}_{\sss (1,1)}}+ {{\hat
M}_{\sss (2,2)}}$, where ${{\hat M}_{\sss (1,1)}}={M_0}{\sigma _{\sss
1}}$ and ${{\hat M}_{\sss (2,2)}}={M}{\bf 1}+{M_3}{\sigma _{\sss
3}}$. Note that $[{{\hat M}_{\sss (1,1)}},{{\hat M}_{\sss (2,2)}}]=0$
has a nontrivial solution if and only if ${M_3}=0$ which corresponds
to the point of unbroken $\discrete$, assuming ${\kappa_{\sss 2}}=0$
in the lagrangian. One can easily verify that these considerations
apply to the mass-squared matrix as well. Therefore, we have the
matching conditions ${{\cal M}^{(+)}_{ab}}\propto{\delta_{ab}}$, and
at the point of enhanced $\discrete$ symmetry, also ${{\cal
M}^{(+)}_{\alpha\beta}}\propto{\delta_{\alpha\beta}}$. At low energies
this gauge theory with vectorlike flavor symmetries gives rise to
hadronic scattering amplitudes that exhibit classic Regge behavior!
One might be tempted to believe that, as in QCD, nonperturbative
effects change this picture. This is unlikely.  Because $G$ is
vectorlike, this gauge theory admits a lattice definition that
respects $G$ {\it at all scales}~\cite{nn}.  Therefore, in contrast
with QCD, the lagrangian of the vectorlike theory should exhibit the
symmetry properties of the nonperturbative regime.

\section{The Vectorlike Sigma Model}

It is easy to construct a sigma model that respects all of the
symmetries of the underlying vectorlike theory~\cite{beane2}. We can
simply ignore the gauge degrees of freedom, identify quarks with
baryons (for $N$ odd) and choose ${\cal M}={F_\pi}+i2{\pi_a}{T_a}$.
The physical baryon states, ${\cal N}=({{\cal N}_+}\;{{\cal
N}_-})^{\sss T}$, are given by

\begin{equation}
\psi = (\sin\phi +i{\sigma _{\sss 2}}\cos\phi ){\cal N}
\label{eq:physstdef}
\end{equation} 
where $\cot 2\phi ={\kappa_{\sss 2}}{F_\pi}/{M_0}$, and have masses:
$M_{\pm}= {\kappa_{\sss 1}}{F_\pi} \pm (\cos 2\phi{\kappa_{\sss
2}}{F_\pi} + \sin 2\phi{M_0})$.  The axial vector coupling matrix is
given by

\begin{equation}
{{\hat g}_{\sss A}}= \left(\matrix{-\cos{2\phi} & -\sin{2\phi} \cr
-\sin{2\phi} & \quad\cos{2\phi} }\right).
\label{eq:gamatrix}
\end{equation} 
Note that $({\hat g}_{\sss A})^2={\bf 1}$, a statement of the
Adler-Weisberger sum rule. Since $Tr\,({\hat g}_{\sss A})=0$,
${\pi_0}\rightarrow\gamma\gamma$ vanishes for all values of $\phi$, as
expected since the underlying theory is anomaly free~\cite{beane2}.

Note the point of enhanced symmetry. If ${\kappa_{\sss 2}}=0$, the
diagonal elements of the axial vector coupling matrix vanish.  This is
because there is a $\discrete$ symmetry if ${\kappa_{\sss 2}}=0$.  We
can then assign multiplicatively conserved $\discrete$ charges to each
physical state.  We assign ${\cal N}_+$ charge $+1$, and ${\cal N}_-$
and $\pi$ charge $-1$.  Clearly the diagonal elements of the
axialvector coupling matrix vanish because they do not respect the
$\discrete$ symmetry. Off-diagonal elements are unity. Here we have
$M={\kappa_{\sss 1}}{F_\pi}$ and ${M_3}={\kappa_{\sss
2}}{F_\pi}{\sigma _{\sss 3}}$ and so it is clear that the baryon
mass-matrix exhibits the claimed symmetry structure, and the point of
enhanced $\discrete$ symmetry, ${\kappa_{\sss 2}}=0$, corresponds to
the constraint $[{{\hat M}_{\sss (1,1)}},{{\hat M}_{\sss (2,2)}}]=0$.
Of course one would find precisely the same ${{\hat g}_{\sss A}}$ by
simply assuming the equivalent Regge inspired sum rules.

\section{Outlook}

The nonlinear realization of a broken symmetry encodes symmetry
information that is unique to all underlying theories with the same
pattern of symmetry breaking (chiral perturbation theory), and
symmetry information that is determined by matching to the underlying
theory (asymptotic matching). We have shown that while a matching
condition which determines the true asymptotic behavior of the total
pion-hadron cross-section is mysterious from the QCD point of view, it
is easy to construct a new asymptotically free gauge theory with QCD
flavor symmetries in which this matching condition is manifest. The
fundamental property of this new gauge theory is that all flavor
symmetries are vectorlike. This gauge theory therefore admits a
lattice formulation that respects QCD-like flavor symmetries and in
turn provides a theoretical laboratory which can rigorously address
the issue of how these flavor symmetries are realized in the vacuum.

\vspace{0.1in}

\noindent This work was supported by the U.S. Department of Energy
grant DE-FG02-93ER-40762 (U.Md.PP\# 99-029). I thank CK Chow for
valuable discussions.

\end{document}